\documentclass[journal]{IEEEtran}

\usepackage{xcolor,soul,framed} 

\colorlet{shadecolor}{yellow}
\usepackage[pdftex]{graphicx}

\graphicspath{{../pdf/}{../jpeg/}}
\DeclareGraphicsExtensions{.pdf,.jpeg,.png}

\usepackage[cmex10]{amsmath}
\usepackage{array}
\usepackage{mdwmath}
\usepackage{mdwtab}
\usepackage{eqparbox}
\usepackage{url}
\usepackage[nolist]{acronym}

\usepackage{cite}
\usepackage{booktabs} 
\usepackage{multirow}
\usepackage{multicol}
\usepackage{amsfonts}
\usepackage{float}

\usepackage{amssymb}
\usepackage{pifont}
\include{acro}

\begin{document}
    \title{Speaker Extraction with Co-Speech Gestures Cue}
    \author{Zexu~Pan, Xinyuan~Qian,  and~Haizhou~Li,~\IEEEmembership{Fellow,~IEEE}

    \thanks{This research is supported by  
    The Science and Engineering Research Council, Agency for Science, Technology and Research (A*STAR), Singapore, through the National Robotics Program under Human-Robot Interaction Phase 1 (Grant No. 192 25 00054), 
    by the Deutsche Forschungsgemeinschaft (DFG, German Research Foundation) under Germany's Excellence Strategy (University Allowance, EXC 2077, University of Bremen), and by the Guangdong Provincial Key Laboratory of Big Data Computing, The Chinese University of Hong Kong, Shenzhen under Grant No. B10120210117-KP02, UDF01002333, and UF02002333.
    (\textit{Corresponding author: Xinyuan Qian)}}
    \thanks{Zexu Pan is with the Integrative Sciences and Engineering Programme, and the Institute of Data Science, National University of Singapore, 119077 Singapore (e-mail: pan\_zexu@u.nus.edu).}
    \thanks{Zexu Pan, and Haizhou Li are with the Department of Electrical and Computer Engineering, National University of Singapore, 119077 Singapore (e-mail: haizhou.li@u.nus.edu).}
    \thanks{Xinyuan Qian and Haizhou Li are with the Guangdong Provincial Key Laboratory of Big Data Computing, the Chinese University of Hong Kong, Shenzhen, 518172 China (e-mail: qianxinyuan@cuhk.edu.cn)}
    \thanks{Haizhou Li is also with the School of Data Science, the Chinese University of Hong Kong, Shenzhen, 518172 China, and the University of Bremen, 28359 Germany.}
    }

\maketitle

\begin{abstract}
Speaker extraction seeks to extract the clean speech of a target speaker from a multi-talker mixture speech. There have been studies to use a pre-recorded speech sample or face image of the target speaker as the speaker cue. In human communication, co-speech gestures that are naturally timed with speech also contribute to speech perception. In this work, we explore the use of co-speech gestures sequence, e.g. hand and body movements, as the speaker cue for speaker extraction, which could be easily obtained from low-resolution video recordings, thus more available than face recordings. We propose two networks using the co-speech gestures cue to perform attentive listening on the target speaker, one that implicitly fuses the co-speech gestures cue in the speaker extraction process, the other performs speech separation first, followed by explicitly using the co-speech gestures cue to associate a separated speech to the target speaker. The experimental results show that the co-speech gestures cue is informative in associating with the target speaker.
\end{abstract}


\IEEEpeerreviewmaketitle

\section{Introduction}
\label{sec:introduction}
\IEEEPARstart{S}{peech} is the most natural way of human communication. However, the performance of computer processing of speech, such as automatic speech recognition~\cite{yue2019end}, speaker localization~\cite{qian2021multi}, active speaker detection~\cite{tao2021someone}, and speech emotion recognition ~\cite{pan2020multi} degrades dramatically in the presence of interfering speakers. This prompts us to study ways to extract speech similar to how humans perceive.

Speech separation seeks to separate a mixture speech into individual clean speech streams by speakers~\cite{hershey2016deep,liu2019divide,luo2019conv,luo2020dual,kolbaek2017multitalker,zeghidour2020wavesplit}. In a cocktail party, humans use the intrinsic ability to attentively listen to a speaker of interest, i.e. target speaker~\cite{bronkhorst2000cocktail,cherry1953some}. The study of speaker extraction seeks to mimic such selective attention and extracts only the speech of the target speaker, from an adverse acoustic environment~\cite{pan2021reentry}. While speaker extraction does not require prior knowledge of the number (no.) of speakers like speech separation, it relies on some reference cues for the extraction.

In speaker extraction, a pre-recorded speech sample may serve well as the reference cue~\cite{Chenglin2020spex,spex_plus2020,delcroix2020improving,wang2019voicefilter,xiao2019single,shi2020speaker}. However, such a pre-recording may not be available always, e.g. when a robot wants to listen to a  passer-by. There have been recent studies where the real-time face video sequence of the target speaker is used as the auxiliary reference~\cite{ephrat2018looking,wu2019time,ochiai2019multimodal,pan2020muse,pan2021usev,koichiro2021,michelsanti2021overview}.  This technique works well when we are able to capture high-resolution lip movements, e.g. during video conferencing, but not in the case when we wear face masks during the COVID-19 pandemic~\cite{ciotti2020covid}.

Embodied human communication encompasses interactions between verbal (speech) and non-verbal (body posture, hand gestures, and head nods) behaviors~\cite{ahuja2020style,kendon2004gesture}. For example, when producing a speech, we move our hands up and down to emphasize or describe the outline of a shape for effective communication. Our auditory perception improves by observing the accompanying gestures of the speaker. Gestures that are time aligned with the verbal and vocal content of communication are known as co-speech gestures~\cite{kendon2011gesticulation}. Co-speech gestures are highly correlated with the speech's conceptual content and prosody, neuroscience studies suggest that co-speech gestures are beneficial for speech production and perception~\cite{wagner2014gesture,obermeier2015speaker}.

As far as the visual cue is concerned, the upper-body is more visible than the lips, especially from a distant view. In this work, we propose a speaker extraction framework that uses the upper-body video recording of the target speaker as the reference cue. It is a departure from either the speech cue or lip motion cue. To the best of our knowledge, this is the first study to employ the co-speech gestures cue in a computational model for speaker extraction. This work is particularly useful when, accompanying the speech, only a low-resolution video recording is available. A related task to our work is using motion or gesture cues for sound source separation~\cite{gan2020music,7951787,zhu2022visually}, our work differs from them by exploring the relationship between co-speech gestures and human speech.

We study two novel neural architectures that perform speaker extraction with the co-speech gestures cue: a) The speaker extraction with co-speech gestures cue (SEG) network that directly extracts the target speech from the mixed speech, by taking co-speech gestures as a reference in the process. b) The cascaded dual-path recurrent neural network (DPRNN)~\cite{luo2020dual} with a gesture-speech recognition (GSR) network named DPRNN-GSR,
that performs speech separation first using DPRNN and explicitly associates a separated speech to the target speaker using GSR. We show that both networks perform well on the `in the wild' YouTube gesture dataset~\cite{yoon2019robots}.

\section{Proposed networks}
\label{sec:methodology}
Let $s(\tau)$ be the target speech and $b_{i}(\tau)$ be the interfering speech, where $i \in \{1,...,I\}$ denotes the index of interfering speakers. The mixture speech $x(\tau)$ is defined as:
\begin{equation}
    \label{eqa:speaker_extraction}
    x(\tau) = s(\tau) + \sum_{i=1}^{I}b_{i}(\tau)
\end{equation}
A speaker extraction algorithm seeks to extract $\hat{s}(\tau)$ that approximates $s(\tau)$ from a $(I+1)$-talker mixture speech\footnote{A variable with subscript ($\tau$) represents time-domain waveform samples, while a variable with subscript ($t$) represents frame-based embeddings in the latent space.}.

\begin{figure}
  \centering
  \includegraphics[width=0.99\linewidth]{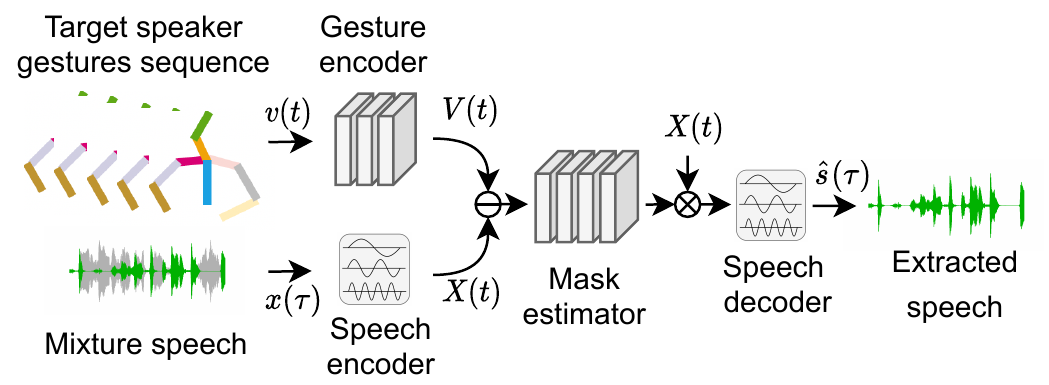}
  \vspace*{-7mm}
  \caption{The proposed speaker extraction with co-speech gestures cue (SEG) network. It implicitly fuses the co-speech gestures cue $V(t)$ with the mixture speech embeddings $X(t)$ to estimate a mask that only lets the target speaker pass. The symbol $\ominus$ represents the concatenation of features along the channel dimension; the symbol $\otimes$ represents element-wise multiplication.}
  \vspace*{-3mm}
  \label{fig:seg}
\end{figure}

\subsection{SEG network}
The universal speaker extraction with visual cue (USEV) network~\cite{pan2021usev} performs speaker extraction conditioning on the lip movements cue during mask estimation for a target speaker. Motivated by this idea, we propose the SEG network, which takes the co-speech gesture cue during the mask estimation. Like the USEV network, the SEG network is invariant to the number of speakers in the mixture speech.

\subsubsection{Architecture}
As shown in Fig.~\ref{fig:seg}, the SEG network consists of four components. a) The speech encoder transforms the time-domain mixture speech waveform samples $x(\tau)$ into mixture speech embeddings $X(t)$. b) The gesture encoder encodes the target speaker gestures sequence $v(t)$ into the co-speech gestures cue $V(t)$. c) The mask estimator takes the concatenated $X(t)$ and $V(t)$ as input and estimates a mask that only lets the target speaker pass from $X(t)$. The mask is then element-wise multiplied with $X(t)$, to produce the masked mixture speech embeddings. d) The speech decoder takes in the masked mixture speech embeddings and transforms them back to time-domain waveform samples, of which the latter is the extracted target speech $\hat{s}(\tau)$.

The proposed SEG network adopts a similar network architecture as the USEV network~\cite{pan2021usev}, except that the SEG network employs the gesture encoder for co-speech gestures, whereas the USEV network employs a visual encoder for the lip movements. The use of co-speech gestures, instead of lip movements, greatly improves the accessibility of the SEG network in practice.

\subsubsection{Gesture encoder}
The input to the gesture encoder $v(t)$ is a sequence of human upper-body poses consisting of the 3-dimensional (3D) coordinates of ten spine-centered joints, i.e. head, neck, nose, spine, left/right (L/R) shoulders, L/R elbows, and L/R wrists. The recurrent neural networks with the bidirectional scheme are effective in modeling the temporal variations of gestures~\cite{chu2014synchronization,yoon2020speech}. We design the gesture encoder as a $N_{ge}$-layered BLSTM, and up-sample the gesture representations at the end of the BLSTM such that $V(t)$ and $X(t)$ have the same temporal resolution.

\subsubsection{Training objective}
We adopt the negative scale-invariant signal-to-noise ratio (SI-SDR)~\cite{le2019sdr} as the loss function to measure the signal quality of the extracted speech:
\begin{equation}
    \mathcal{L}_{SI\mbox{-}SDR} = - 10 \log_{10} ( \frac{||\frac{<\hat{s},s>s}{||s||^2}||^2}{||\hat{s} - \frac{<\hat{s},s>s}{||s||^2}||^2}),
    \label{eqa:sisdr}
\end{equation}
which is applied to the output of the SEG network between the extracted speech $\hat{s}(\tau)$ and the clean speech $s(\tau)$. We omit the subscript ($\tau$) in the loss functions for brevity. 

\begin{figure*}
    \begin{minipage}[b]{.32\linewidth}
      \centering
      \centerline{\includegraphics[width=\linewidth]{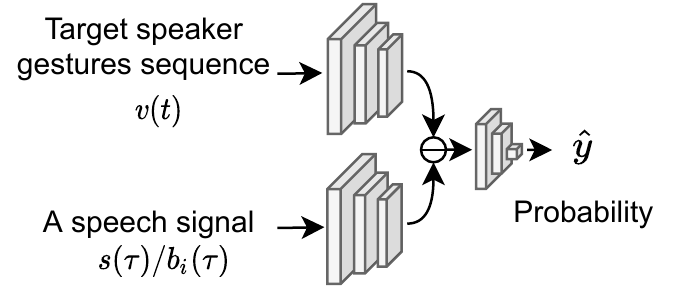}}
      \centerline{\scalebox{0.8}{(a) Gesture-speech recognition (GSR) network.}}\medskip
      \vspace*{-3mm}
    \end{minipage}
    \hfill
    \begin{minipage}[b]{0.68\linewidth}
      \centering
      \centerline{\includegraphics[width=\linewidth]{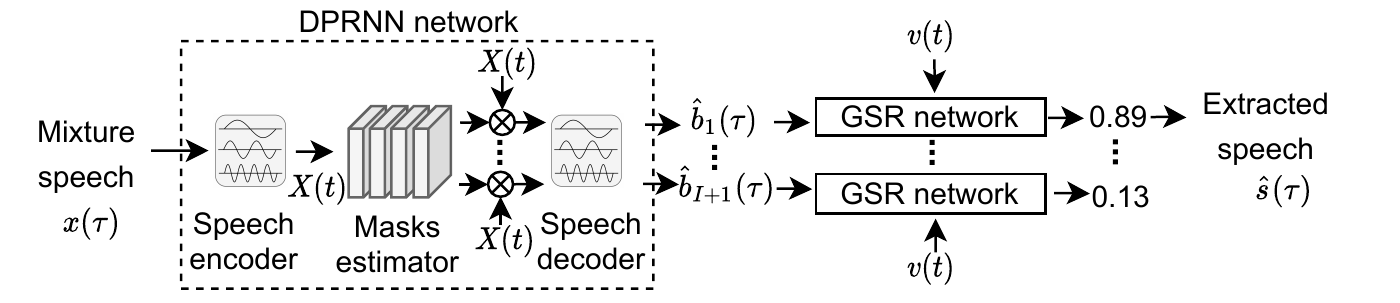}}
      \centerline{\scalebox{0.8}{(b) DPRNN-GSR network} }\medskip
      \vspace*{-3mm}
    \end{minipage}
    \caption{In (a), the proposed gesture-speech recognition network. It outputs the probability of the given co-speech gestures sequence and the speech being paired from the same video. In (b), the proposed DPRNN-GSR network for target speaker extraction, that consists of the cascaded DPRNN network and the GSR network. It first performs speech separation with the DPRNN network, followed by the GSR network to recognize which of the separated speech utterance belongs to the target speaker.}
    \label{fig:dprnn-gsr}
    \vspace*{-3mm}
\end{figure*}

\subsection{DPRNN-GSR network}
In the case when we know in advance the number of speakers in a mixture speech, without any reference cue, the DPRNN~\cite{luo2020dual} network performs very well for speech separation. We propose to form a cascaded network, named DPRNN-GSR network as shown in Fig.~\ref{fig:dprnn-gsr} (b), that performs speech separation first with a DPRNN network, which is followed by a GSR network to identify which of the output speech streams from the DPRNN network aligns well with the co-speech gestures of the target speaker. 

\subsubsection{DPRNN network}
The DPRNN~\cite{luo2020dual} network is illustrated in the dotted box in Fig.~\ref{fig:dprnn-gsr} (b). The mask estimator of the DPRNN network produces a mask for every speaker, and the speech decoder reconstructs the time-domain speech waveform $\hat{b}_j(\tau)$, $j \in \{1,..., I+1\}$ from every masked speech embeddings. The DPRNN network is trained with the negative SI-SDR as the loss function as shown in Eq.~\ref{eqa:sisdr}, the utterance-level permutation invariant training (PIT) is used to address the output permutation problem~\cite{kolbaek2017multitalker}.

\subsubsection{GSR network}
As shown in Fig.~\ref{fig:dprnn-gsr} (a), the GSR network is introduced to select a $\hat{b}_j(\tau)$ that best matches the target speech $s(\tau)$, given the target co-speech gestures sequence. 

Self-supervised learning for audio-visual synchronization detection has been well studied ~\cite{chung2016out,chung2019perfect,afouras2020self,owens2018audio,pan2021reentry}, where the positive samples are the paired audio and visual signals, and the negative samples are the unpaired or temporally misaligned audio and visual signals. Inspired by the SLSyn network~\cite{pan2021reentry}, which is effective at detecting the synchronization between a speech and a lip movement, we propose the GSR network that has a similar network architecture as the SLSyn network.

The GSR network takes a speech and a co-speech gestures sequence as the inputs, and outputs the probability of the speech and co-speech gestures sequence being paired. The paired samples are the target co-speech gestures sequence and the target clean speech (e.g. $v(t)$ and $s(\tau)$), while the unpaired samples are the target co-speech gestures sequence and the interfering clean speech, respectively (e.g. $v(t)$ and $b_i(\tau)$).

We minimize the following binary cross-entropy loss for the GSR network training:
\begin{equation}
    \mathcal{L}_{BCE} = - y\log(\hat{y}) - (1-y)\log(1-\hat{y}),
\end{equation}
where $y \in \{0,1\}$ indicates whether the speech and co-speech gestures are paired, while $\hat{y}$ is the predicted probability.

\subsubsection{Training and inference}
During training, the DPRNN network and the GSR network are trained independently. During inference, the two networks are cascaded, the DPRNN first outputs $I+1$ separated speech utterances, each of the separated speech utterances is then passed through the GSR network together with $v(t)$, to output a probability score. The $b_j(\tau)$ with the highest probability score is selected as the extracted speech $\hat{s}(\tau)$ and the rest are discarded.

The GSR network is trained with clean speech utterances as input, while the separated speech utterances of DPRNN are used during inference. To our pleasant surprise, the DPRNN performs exceptionally well. The experimental results suggest that this mismatch between training and inference is negligible.

\begin{table*}
    \centering
    \caption{The performance of speech separation and our proposed speaker extraction networks.}
    \vspace*{-2mm}
    \begin{tabular}{c c c c c c c c} 
       \toprule
       Dataset  &Task   &Network          &SI-SDRi (dB)   &SDRi (dB)  &PESQi      &STOIi      &Accuracy (\%) \\
       \midrule
       \multirow{5}*{YGD-2mix}
       &Speech separation
                        &DPRNN~\cite{luo2020dual} (PIT association)         &14.51          &14.75      &0.983      &0.209      &-    \\
       \cmidrule{2-8}
       &\multirow{3}*{Speaker extraction}  
                        &DPRNN~\cite{luo2020dual} (random association)          &-6.81          &-1.80      &-0.025      &-0.168     &51.10    \\
                        &&Our DPRNN-GSR     &6.19           &8.05       &0.585      &0.058      &80.77  \\        
                        &&Our SEG          &9.12           &10.04       &0.789      &0.111      &86.13 \\
       \midrule
       \multirow{5}*{YGD-3mix}
       &Speech separation
                        &DPRNN~\cite{luo2020dual} (PIT association)          &11.64          &12.00      &0.779      &0.223      &-    \\
       \cmidrule{2-8}
       &\multirow{3}*{Speaker extraction}  
                        &DPRNN~\cite{luo2020dual} (random association)         &-4.33          &-1.73      &0.001      &-0.098     &45.33    \\
                        &&Our DPRNN-GSR    &1.78            &3.53       &0.309      &0.022      &65.30  \\        
                        &&Our SEG         &4.95            &5.34       &0.342      &0.089      &76.47 \\
       \bottomrule
    \end{tabular}
    \vspace*{-3mm}
    \label{table:result}
\end{table*}

\begin{figure*}[t]
\begin{minipage}[t]{.23\linewidth}
  \centering
  \centerline{\includegraphics[width=\linewidth]{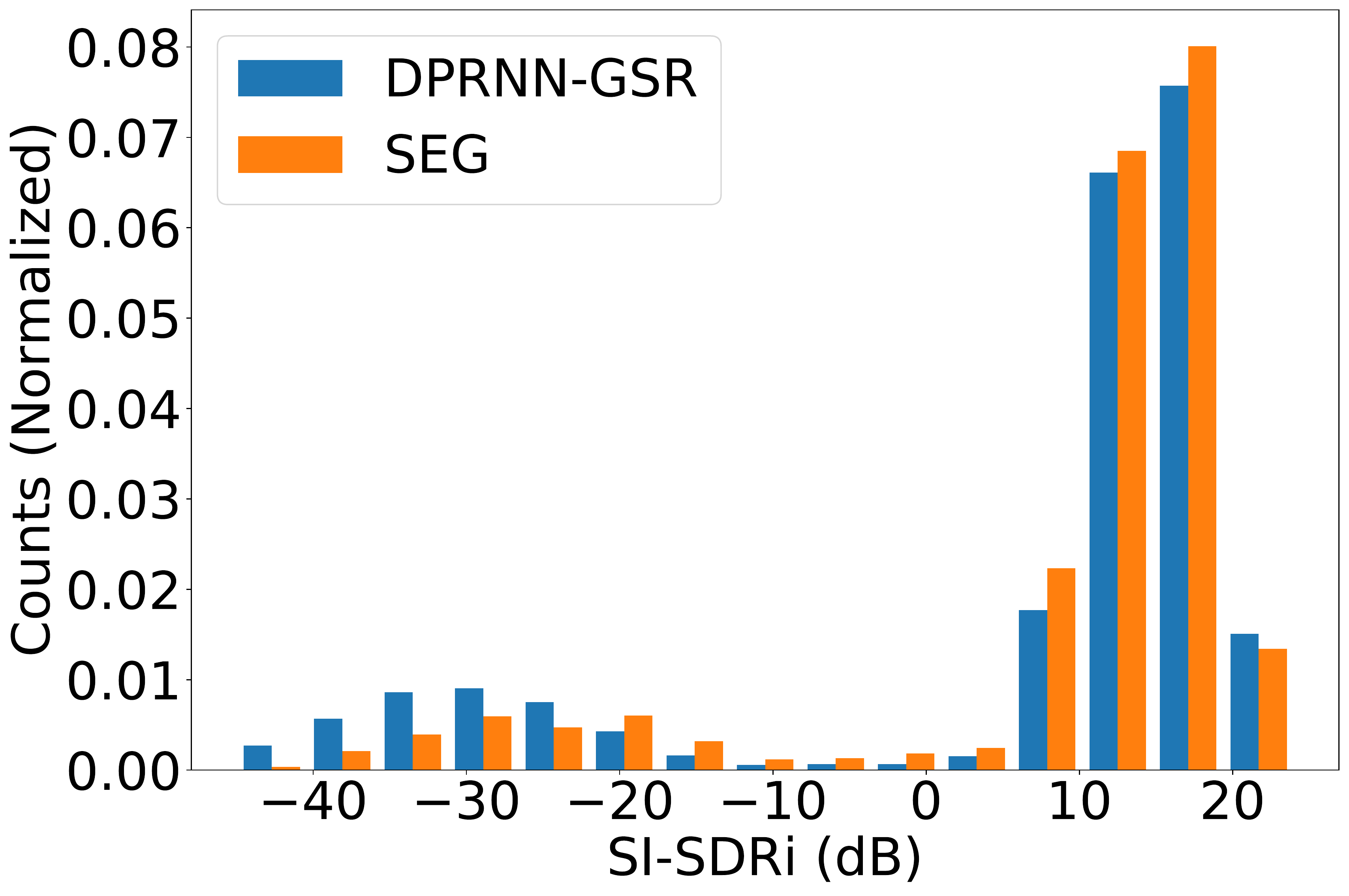}}
  \vspace*{-2mm}
  \caption{The histogram of SI-SDRi by DPRNN-GSR and SEG networks.}\medskip
  \label{fig:count}
\end{minipage}
\hfill
\begin{minipage}[t]{.23\linewidth}
  \centering
  \centerline{\includegraphics[width=\linewidth]{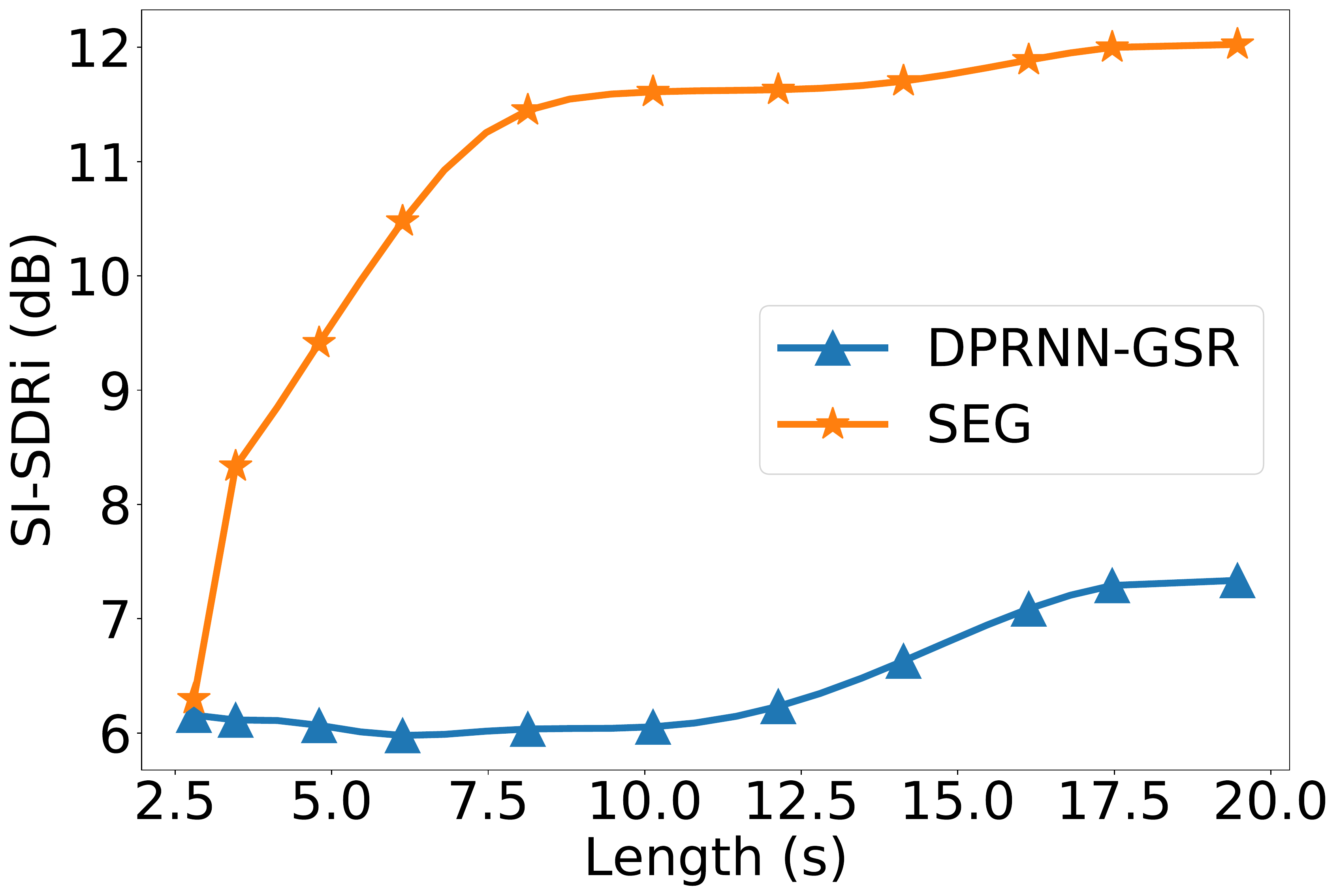}}
  \vspace*{-2mm}
  \caption{The average SI-SDRi by DPRNN-GSR and SEG networks against the utterance length. }\medskip
  \label{fig:time}
\end{minipage}
\hfill
\begin{minipage}[t]{.23\linewidth}
  \centering
  \centerline{\includegraphics[width=\linewidth]{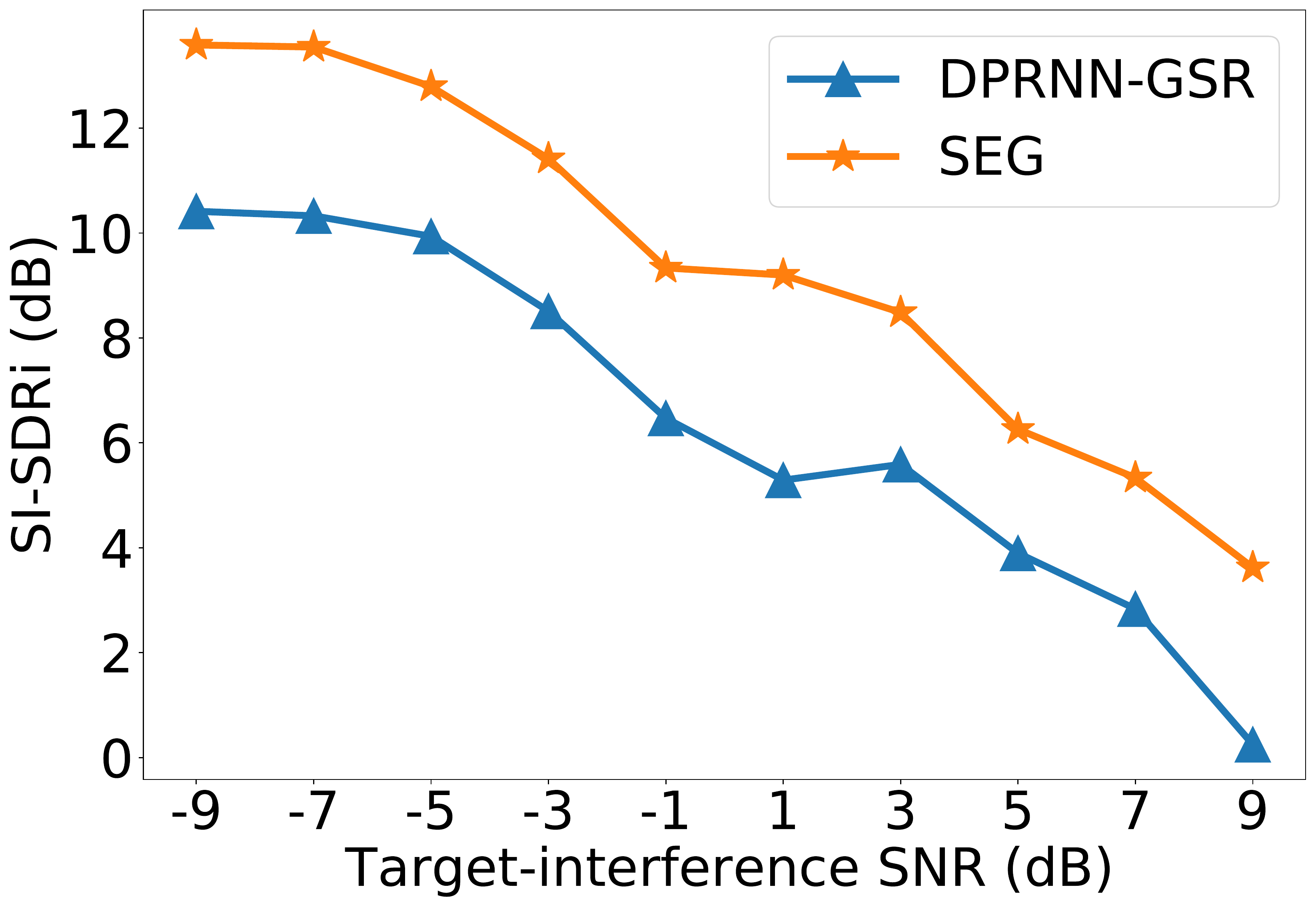}}
  \vspace*{-2mm}
  \caption{The average SI-SDRi by DPRNN-GSR and SEG networks against the target-interference SNR. }\medskip
  \label{fig:sisdr_snr}
\end{minipage}
\hfill
\begin{minipage}[t]{.23\linewidth}
  \centering
  \centerline{\includegraphics[width=\linewidth]{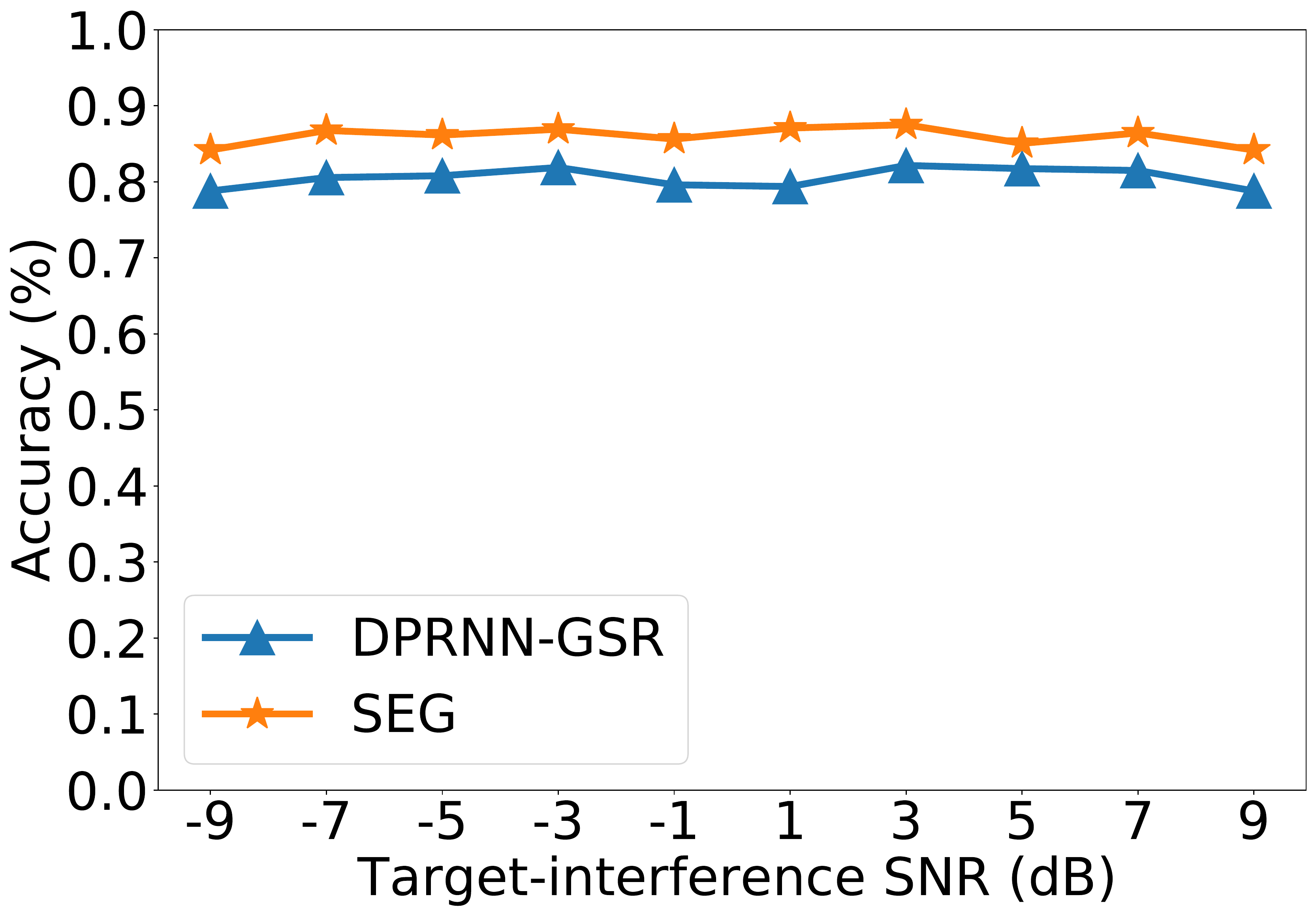}}
  \vspace*{-2mm}
  \caption{The average accuracy by DPRNN-GSR and SEG networks against the target-interference SNR. }\medskip
  \label{fig:accuracy}
\end{minipage}
\vspace*{-5mm}
\end{figure*}

\section{Experimental setup}
\label{sec:experiment}
\subsection{Dataset}
We simulate a 2-speaker mixture speech (YGD-2mix) dataset and a 3-speaker mixture speech (YGD-3mix) dataset to evaluate our proposed speaker extraction networks using a YouTube gesture dataset~\cite{yoon2019robots,yoon2020speech}. The YouTube gesture dataset consists of 1,696 TED videos. The train, validation, and test sets consist of 27,611, 3,654, and 3,475 video segments respectively.

The 2D poses sequences of the speakers in the videos are provided in the YouTube gesture dataset. According to~\cite{yoon2020speech}, the sequences of 2D poses are further converted to sequences of 3D poses by using the 3D pose estimator~\cite{pavllo20193d}. The 3D poses are sampled at 15 frames per second, and are used as the co-speech gestures sequence in this work.

We simulate 200,000, 5,000, and 3,000 mixture speech utterances to form the train, validation, and test sets, respectively, for the YGD-2mix dataset and the YGD-3mix dataset. Each interfering speech is mixed with the target speech at a random Signal-to-Noise ratio (SNR) set between 10 and -10 dB. The longer speech is truncated to the length of the shorter one. We simulate more short utterances than long utterances as short utterances are considered harder. The audios are sampled at 16 kHz. Speakers in different sets do not overlap, which allows us to perform speaker-independent evaluations\footnote{The code is available at https://github.com/zexupan/seg}.


\subsection{Model configuration}
The hyper-parameters of the SEG network follow the USEV network~\cite{pan2021usev} except for the gesture encoder. In the gesture encoder, the $N_{ge}$ is set to 5, the hidden size and dropout probability of the BLSTM are set to 128 and 0.3.

In the DPRNN-GSR network, the hyper-parameters of the DPRNN network follow the original implementation~\cite{luo2020dual}, except that the kernel size of the convolutional layer in the speech encoder is set to 40. The GSR network has hyper-parameters similar to the SLSyn network~\cite{pan2021reentry}.

\subsection{Training}
For the SEG network training, the Adam optimizer is used with an initial learning rate set to 0.0005. We half the learning rate if the best validation loss (BVL) does not decrease for 6 epochs, and stop the training if the BVL does not decrease for 10 epochs. The YGD-2mix dataset and the YGD-3-mix dataset are used, where each data tuple consists of $\langle x(\tau), v(t), s(\tau) \rangle$.

For the DPRNN network training, the Adam optimizer is used with an initial learning rate set to 0.001. We half the learning rate if the BVL does not decrease for 6 epochs, and stop the training if the BVL does not decrease for 10 epochs. The YGD-2mix dataset and the YGD-3mix dataset without gestures are used, where each data tuple consists of $\langle x(\tau), s(t), b_{1}(\tau), ... , b_{I}(\tau)\rangle$.

For the GSR network training, the Adam optimizer is used with an initial learning rate set to 0.0001. The learning rate decreases by 10\% after every training epoch, and training stops if the BVL does not decrease for 5 epochs. The original YouTube gesture dataset is used, where each data tuple consists of $\langle v(t), s(\tau)/b_{i}(\tau)\rangle$, where $i \in \{1,...,I\}$.

\section{Result}
\label{sec:results}

\subsection{Gesture-speech recognition}
We first study the correlation between a speech sample and its accompanying co-speech gestures. We report the performance of the GSR network under three different settings: 1) The GSR network achieves 76.07\% accuracy, in verifying whether a clean speech matches with a given co-speech gestures sequence; 2) The GSR network achieves 82.17\% accuracy, in selecting one out of two clean speech that matches a given co-speech gestures sequence; 3) The GSR network achieves 70.77\% accuracy, in selecting one out of three clean speech that matches a given co-speech gestures sequence. The results suggest that there is a strong correlation between a speech and its co-speech gestures, which supports the hypothesis that we could use co-speech gestures as the reference cue for target speaker extraction.
 
\subsection{Target speaker extraction}
We perform speaker extraction experiments on the YGD-2mix and YGD-3mix datasets, and report the results in Table~\ref{table:result}. We report the SI-SDR improvement (SI-SDRi) and the signal-to-noise ratio improvement (SDRi) to measure the extraction quality, we also report the perceptual evaluation of speech quality improvement (PESQi) and the short-term objective intelligibility improvement (STOIi) to measure the perceptual quality and the intelligibility. The improvements are calculated with respect to the mixture speech. If an extracted speech utterance has a positive SI-SDRi, we consider that the network has correctly extracted the speech for the target speaker. We define the accuracy of target speaker extraction as the ratio of the no. of correctly extracted speech utterances to the total no. of test utterances. The higher the better for all metrics.

It is observed that the DPRNN network performs remarkably well for speech separation in terms of signal quality metrics, i.e. SI-SDRi, SDRi, PESQi, and STOIi, where we do not need to identify a target speaker. However, in speaker extraction,  its signal quality performance drops as the performance of the GSR network plays a role, and the accuracy of association between the target speaker identity and the output speech stream matters. If we randomly associate an output speech stream with the target speaker, the DPRNN network will see an accuracy of 50.10\% on the YGD-2mix dataset and 45.33\% on the YGD-3mix dataset.

For the YGD-2mix and YGD-3mix datasets, if the DPRNN network is cascaded with a GSR network to select the target speech with the co-speech gestures cue (DPRNN-GSR), all metrics improve significantly compared to the DPRNN network on the speaker extraction task. The SEG network outperforms the DPRNN-GSR network for the speaker extraction task and achieves the best results for all metrics.

In Fig.~\ref{fig:count}, we show the histogram of SI-SDRi for the YGD-2mix test set samples by the DPRNN-GSR and SEG networks. It is seen that the SEG network has more test samples with positive SI-SDRi. The test samples are distributed at the far two ends, either with a very positive SI-SDRi (extracting the correct target speaker) or a very negative SI-SDRi (extracting the wrong target speaker).

In Fig.~\ref{fig:time}, we show the average SI-SDRi for the YGD-2mix test set samples as a function of the input utterance length. It is seen that as the speech duration increases, the SEG network performs better. This is because a longer duration of co-speech gestures is more informative than a shorter one. The performance of the DPRNN-GSR network remains relatively flat when the utterance length is below 10 seconds, this may be because the performance of the GSR network is less affected in this region.

In Fig.~\ref{fig:sisdr_snr}, we show the average SI-SDRi for the YGD-2mix test set samples with various target-interference SNR. The target-interference SNR is defined as the energy contrast between the target speaker and the interfering speaker in the mixture speech in terms of SNR. As the input mixture becomes less noisy, i.e., with a higher SNR, the SI-SDRi becomes smaller. The SEG network outperforms the DPRNN-GSR network on various target-interference SNR.

In Fig.~\ref{fig:accuracy}, we show the average accuracy for the YGD-2mix test set samples with various target-interference SNR. It is seen that the accuracy is not affected by the target-interference SNR for both networks.

\section{Conclusion}
\label{sec:conclusion}
In this work, we explore the use of the co-speech gestures cue for target speaker extraction. This work is particularly useful when the target speaker is only visible from a distant view. We propose two networks that make different use of the co-speech gestures cue, namely the DPRNN-GSR network and the SEG network. Experimental results show that the co-speech gestures that are highly correlated with the speech signal are very informative in disentangling the target speech from the mixture speech.

\bibliographystyle{IEEEtran}
\bibliography{IEEEabrv,Bibliography}

\end{document}